\documentclass[%
 reprint,
 amsmath,amssymb,
prb,
]{revtex4-1}

\usepackage{graphicx}
\usepackage{dcolumn}
\usepackage{bm}
\usepackage{xcolor}

\begin{document}

\preprint{APS/123-QED}

\title{Non-Perturbative Electron Counting Statistics}

\author{Richard Stones}
\email{r.stones@ucl.ac.uk}
\author{Alexandra Olaya-Castro}%
 \email{a.olaya@ucl.ac.uk}
\affiliation{%
 Department of Physics and Astronomy, University College London, Gower Street, London, WC1E 
}%



\date{\today}

\begin{abstract}
The theory of full counting statistics allows complete characterization of charge transport processes through nanoscale systems. The majority of existing theoretical treatments used to obtain the current cumulants rely on perturbative approximations with respect to either the system-bath coupling or the electronic tunnelling coupling within the system. This is not generally well suited to electron transfer through organic molecules where these couplings can be on the same order of magnitude. Here we present a non-perturbative approach that uses the formalism of  the hierarchical equations of motion to describe the system-bath dynamics while retaining a weak coupling with respect to source and drain leads. The scope of this new framework is demonstrated by comparisons with a perturbative approach for a dimer system coupled to a thermal  bath. The inadequacy of the perturbative approach to describe current fluctuations in a variety of regimes is shown.
\end{abstract}

\pacs{Valid PACS appear here}
\maketitle


Molecular electron transport is a process fundamental to life, being involved in key stages of both photosynthesis and respiration \cite{Gray1996}. Incorporating DNA strands \cite{Fink1999} and individual protein units \cite{Gerster2012} into electric circuits has opened up the possibility of investigating electron transport through organic molecules by analysing the current passing through the system. There is also great potential to use molecular electron transport in novel nanoscale electronic devices \cite{Tao2006,Selzer2006}. The use of single molecule junctions in diodes, switches and transistors can generate a much richer behaviour than similar semiconductor based devices in part due to the strong interaction between electronic and vibrational components of the molecules \cite{Tao2006}. This poses a challenging theoretical problem as the electron-vibrational interaction strength must be treated on the same level as the electronic tunnelling coupling in order to accurately describe electron transport. 

The full counting statistics of electron transport provides dynamical information about electron correlations encoded in the probability distribution of the number of charges transferred through the system \cite{Levitov1993,Blanter2000,Marcos2010}. Features of the electron dynamics such as quantum coherence \cite{Kießlich2007} or the system energy level structure \cite{Belzig2005} can be probed with such a technique, while signatures of the vibrational environment may also be present in the current statistics \cite{Haupt2006}. Existing frameworks which provide access to the full counting statistics usually involve a perturbative description of either the electron-vibrational coupling or the electronic tunnelling coupling within the system \cite{Brandes2005}. In non-Markovian approaches \cite{Braggio2006,Flindt2008,Flindt2010} the current cumulants can be derived from a generalized master equation where the system-environment coupling is often not treated exactly and so the parameter regimes in which they are valid are restricted. Some frameworks do allow for arbitrary system-bath coupling while still considering inexact bath dynamics \cite{Braggio2009}. However, non-perturbative frameworks incorporating an exact bath dynamics have been developed for the full counting statistics of bath observables \cite{Cerillo2016}.

In the following we report a non-perturbative approach to electron counting statistics which provides an exact treatment of the coupling between the system and environment as well as the electronic tunnelling coupling within the system. Using the hierarchical equations of motion \cite{Tanimura2006,Ishizaki2009,Tanimura2012,Kreisbeck2011} we derive an expression for the zero-frequency current cumulants for non-Markovian quantum dissipative systems which has an analogous formulation in the well studied Markovian regime \cite{Flindt2005,Marcos2010,Flindt2010}. We demonstrate the wide range of applicability of our approach by comparison with a perturbative framework.

\begin{figure}[!t]
\centering
\includegraphics[scale=1,trim={1.5cm 0.5cm 1.5cm 0.5cm}]{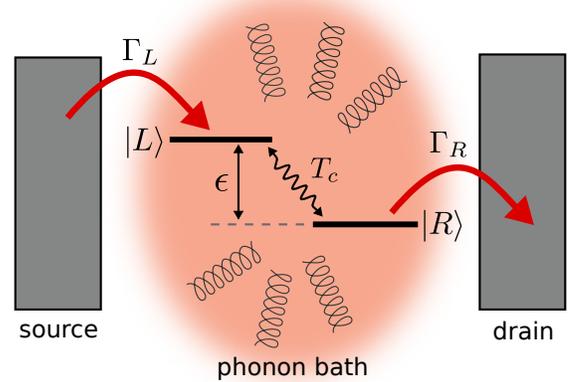}
\caption{Schematic representation of a typical electron counting statistics set up for a dimer system connected to source and drain leads. The dimer consists of discrete electronic states coupled with a harmonic phonon bath. The 'empty' electronic state $|0\rangle$ is not shown.}
\label{dimer-counting-statistics-setup}
\end{figure}

A representation of the model we use is shown in Fig. \ref{dimer-counting-statistics-setup}. We consider a dimer system coupled to fermionic leads which either supply or take away electrons from the system. The molecular dimer consists of discrete electronic energy levels coupled to a bosonic phonon bath. The Hamiltonian for the full system is $H = H_S + H_B + H_L + H_{SB} + H_{SL}$ where
\begin{eqnarray}
H_S &=& \frac{\epsilon}{2} (|L\rangle\langle L|-|R\rangle\langle R|) +  T_c(|L\rangle\langle R| + |R\rangle\langle L|) \\
H_B &=& \sum_q \omega_q b^{\dagger}_q b_q \\
H_L &=& \sum_{k,j=L,R} \epsilon_{j,k} c^{\dagger}_{j,k} c_{j,k}
\end{eqnarray}
are the Hamiltonians for the system, bath and leads respectively, and we have set $\hbar =1$. Here, $\epsilon$ is the energy bias between the left $|L\rangle$ and right $|R\rangle$ dimer sites and $T_c$ is the electronic tunnelling coupling between the sites. $\omega_q$ is the frequency of the qth bosonic mode while $b^\dagger_q$ and $b_q$ create and annihilate a phonon in the qth mode respectively. Finally $\epsilon_{j,k}$ is the energy of an electron in the kth energy level of lead $j$ while $c^\dagger_{j,k}$ and $c_{j,k}$ create and annihilate an electron in the kth energy level of lead $j$ respectively. The system-bath and system-lead coupling are given by
\begin{eqnarray}
H_{SB} &=& \sum_{q} (g_{L,q} |L\rangle\langle L| + g_{R,q} |R\rangle\langle R|) (b^{\dagger}_q + b_q) \\
H_{SL} &=& \sum_{k,j=L,R} v_{j,k}(c^{\dagger}_{j,k}s_j + c_{j,k}s^{\dagger}_j)
\end{eqnarray}
The electron-phonon coupling between site $j$ and phonon mode $q$ is given by $g_{j,q}$ while $\nu_{j,k}$ is the coupling between site $j$ and kth energy level in lead $j$. The operators $s^{\dagger}_j$ and $s_j$ create or annihilate an electron on site $j$ respectively. The coupling between the dimer and the  leads is assumed to be weak such that a second order perturbative expansion can be carried out along with the Born-Markov approximation to obtain the action of the leads on the reduced system \cite{Brandes2005,Harbola2006}. A further assumption that the electronic energy levels lie well within the voltage bias window between the leads is made so the infinite bias limit can be applied \cite{Harbola2006}. This gives the equation of motion for the system and bath in the Schr\"{o}dinger picture
\begin{equation} \label{sys-bath-liouville-eqn-schrodinger-pic}
\frac{d}{dt} \chi(t) = -i[ H_S + H_B + H_{SB}, \chi(t) ] + \mathcal{D}_L\big(\chi(t)\big)
\end{equation}
where $\chi(t) = tr_L \{ \rho_{SB}(t) \otimes \rho_L(t)\}$. The dissipator $\mathcal{D}_L\big(\chi(t)\big)$ derived in Ref. \cite{Harbola2006} describes the action of the leads on the electronic system and has Lindblad form
\begin{equation}
\mathcal{D}_L\big(\chi(t)\big) = \sum_{j=L,R} \Gamma_j \Big[ s_j \rho_S s_j^{\dagger} - \frac{1}{2} \{ s_j^{\dagger}s_j, \rho_S \} \Big].
\end{equation}
The Lindblad operators $s_L = |L\rangle\langle 0|$ and $s_R = |0\rangle\langle R|$ cause transitions between the states where an excess electron occupies either the left or right molecule of the dimer and the empty state of the system $|0\rangle$ where there are no excess charges. The Coulomb blockade is enforced by restricting the electronic subspace such that the identity is $I_S = |0\rangle\langle 0| + |L\rangle\langle L| + |R\rangle\langle R|$.

The hierarchical equations of motion can then be derived with respect to the system-bath coupling \cite{Tanimura2006,Ishizaki2009,Kreisbeck2011,Tanimura2012}. A hierarchy of auxiliary density operators $\sigma^{(\mathbf{n})}(t)$ is formed which keep track of the system-bath correlations \cite{Zhu2012}. The equations of motion are
\begin{eqnarray} \label{heom}
\dot{\sigma}^{(\mathbf{n})}(t) = &-&(i\mathcal{L} + \sum_{j,k} n_{j,k} \nu_{j,k} + \Xi) \sigma^{(\mathbf{n})}(t) \nonumber \\ 
&-&i \sum_{j,k} V^\times_j \sigma^{(\mathbf{n}+1)}(t) \nonumber \\
&-&i \sum_{j,k} n_{j,k} (c_{j,k} V_j)^\times \sigma^{(\mathbf{n}-1)}(t),
\end{eqnarray}
where the electronic system density matrix is given by $\rho_S(t) = \sigma^{(0)}(t)$. The auxiliary density matrices are labelled by the multi-index $\mathbf{n} = \{ n_{L,0}, ... n_{L,K}, n_{R,0}, ... n_{R,K} \}$ while coupling to higher and lower tiers is denoted by $\mathbf{n}\pm1 = \{ n_{L,0}, ... n_{j,k}\pm1, ..., n_{R,K} \}$. The diagonal terms contain the Liouvillian $\mathcal{L}\sigma^{(\mathbf{n})}(t) = -i[H_S, \sigma^{(\mathbf{n})}(t)] + \mathcal{D}_L\big(\sigma^{(\mathbf{n})}(t)\big)$ which takes into account the unitary evolution of the auxiliary density matrices and the interaction with the leads. In Eq. \eqref{heom}, $j$ labels sites $|L\rangle$ and $|R\rangle$ coupled to the phonon bath while $k$ labels the Matsubara terms decaying with frequency $\nu_k$ in the expansion of the bath correlation function \cite{Shi2009}. The coefficients of the correlation function $c_{j,k}$ depend on the form of the spectral density. In practice the hierarchy is truncated at a suitable level that ensures converged results while still enabling tractable numerical computation. Convergence can be improved through appropriate scaling of the auxiliary density matrices \cite{Shi2009} as well as through using a Markovian truncation term \cite{Ishizaki2005} $\Xi = \sum_j \sum_{k=K}^\infty \frac{c_{jk}}{\nu_k}V^\times_j V^\times_j$ to reduce the number of Matsubara terms $K$ needed for convergence, thereby allowing investigation of quantum dynamics at low temperatures.

The hierarchical equations of motion can be formulated as a time-local master equation for the vector of auxiliary density matrices $\sigma(t)$
\begin{equation}
\dot{\sigma}(t) = \mathcal{H}\sigma(t)
\end{equation}
with the hierarchy matrix $\mathcal{H} = \mathcal{H}_0 + \mathcal{H}_J$ acting as a time propagator. $\mathcal{H}_0$ describes the evolution of the system and auxiliary density matrices between electron jump events from the system to the drain lead while $\mathcal{H}_J$ describes these jump events and consists of the part of the dissipator $\mathcal{D}_L\big(\rho_S(t)\big)$ containing $s_R$ in the top level of the hierarchy only.

We consider the current statistics between the system and the drain lead by augmenting $\mathcal{H}_J$ with a counting field $\chi$. The derivation of the counting statistics then follows in analogy with well known Markovian formalisms \cite{Bagrets2003,Brandes2005,Flindt2005,Marcos2010,Flindt2010}. For ease of numerical computation we use the recursive scheme presented in Ref. \cite{Flindt2010}. Introducing a perturbed time propagator $\mathcal{H}(\chi) = \mathcal{H} + \Delta\mathcal{H}(\chi)$ where $\Delta\mathcal{H}(\chi) = (e^{i\chi}-1)\mathcal{H}_J$, we obtain the $\chi$-dependent eigenvalue equation
\begin{equation}
\mathcal{H}(\chi)|0(\chi)\rangle\rangle = \lambda_0(\chi)|0(\chi)\rangle\rangle.
\end{equation}
$|0(\chi)\rangle\rangle$ is the left steady state eigenvector of the hierarchy matrix and $\lambda_0(\chi)$ is the associated eigenvalue which tends to zero as $\chi\rightarrow 0$. The zero-frequency counting statistics are completely determined by this eigenvalue which can be expressed
\begin{equation} \label{eigenvalue}
 \lambda_0(\chi) = \langle\langle\tilde{0}|\Delta\mathcal{H}(\chi)|0\rangle\rangle
\end{equation}
where $\langle\langle\tilde{0}|$ is the right steady state eigenvector of the hierarchy matrix.

\begin{figure*}[t]
\centering
\includegraphics[scale=0.45,trim={1.5cm 0.5cm 1.5cm 1cm}]{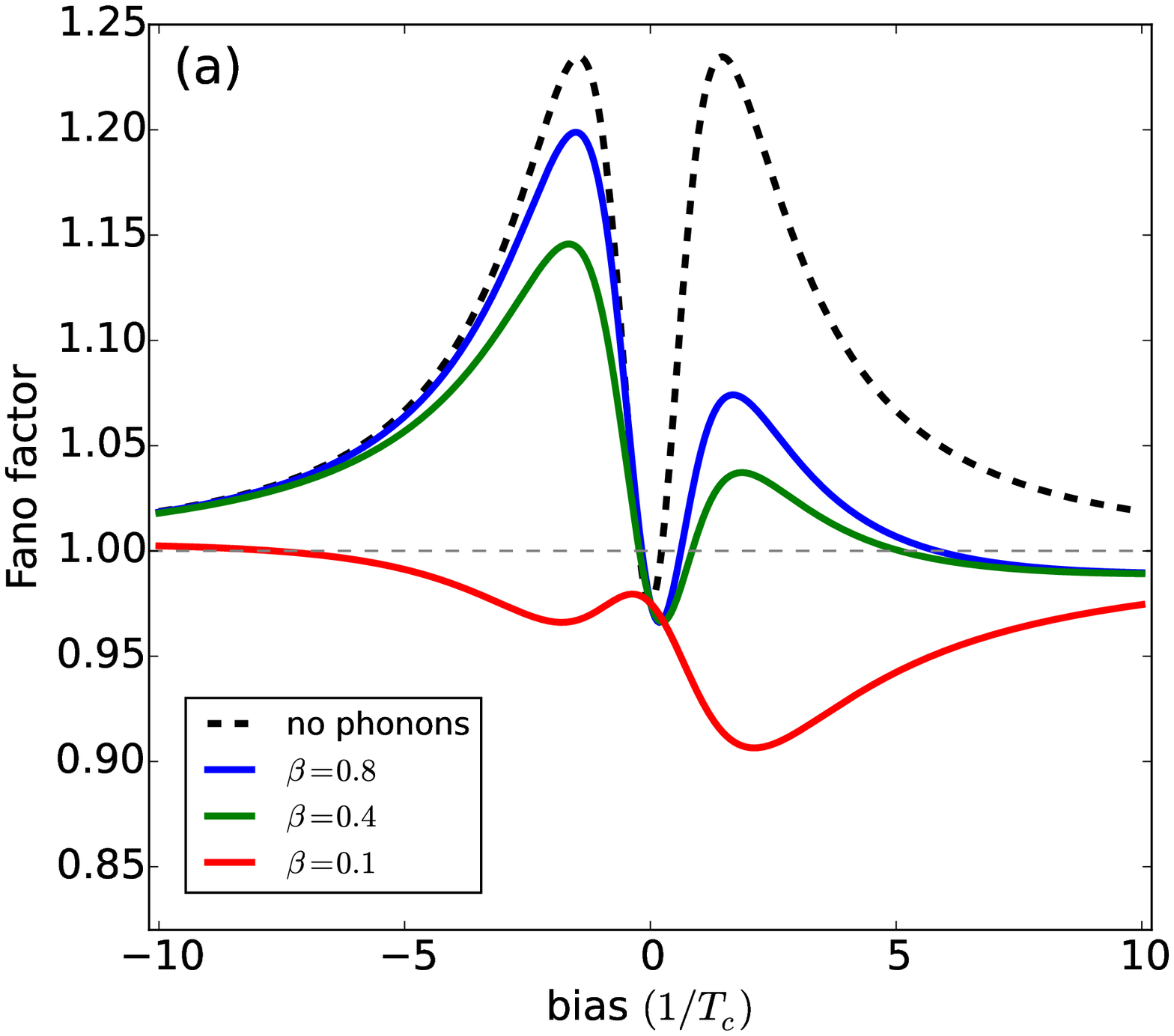}
\hspace{1cm}
\includegraphics[scale=0.45,trim={1.5cm 0.5cm 1.5cm -0.5cm}]{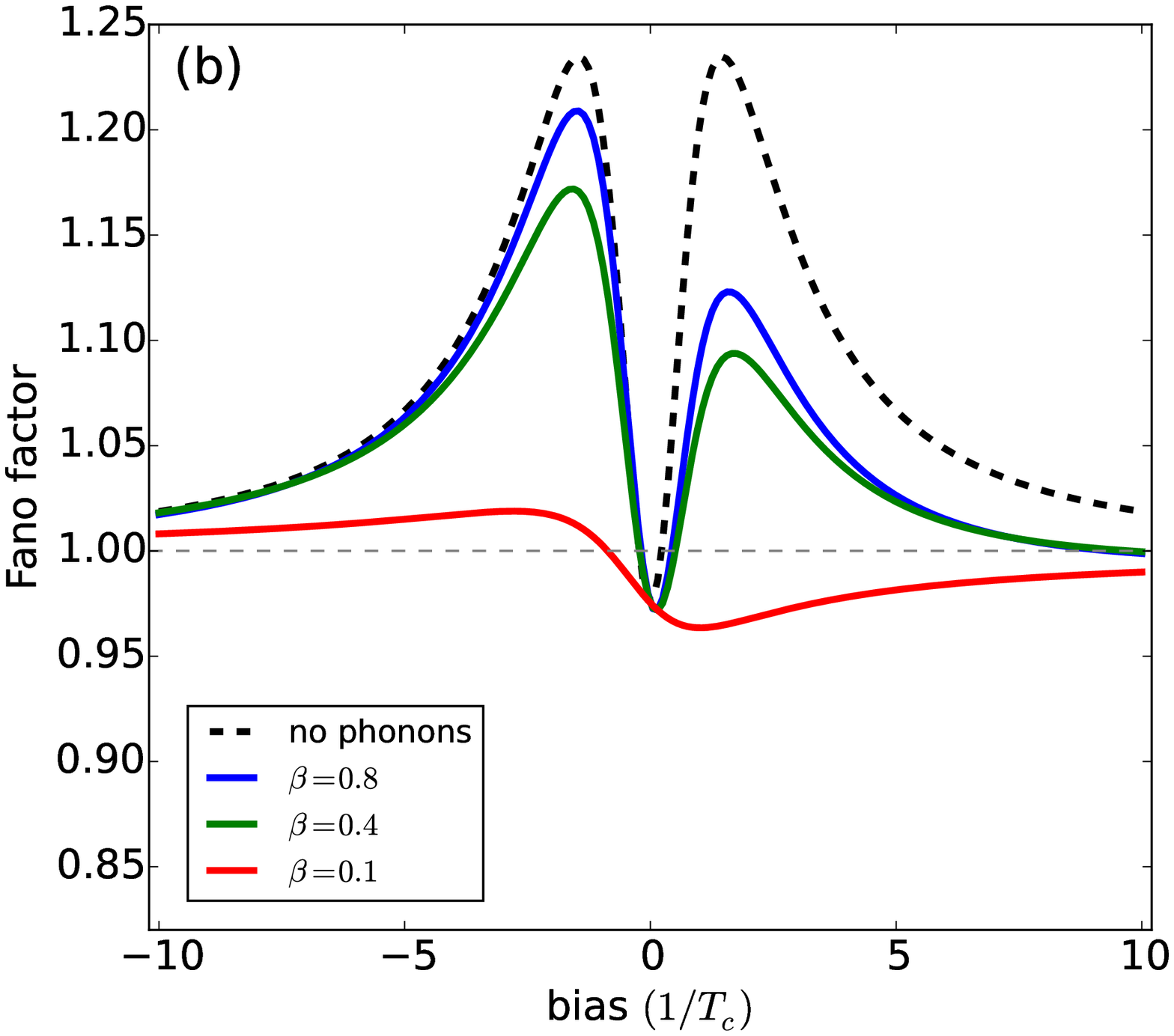}
\caption{Fano factor as a function of dimer energy bias $\epsilon$ calculated using (a) the weak coupling approximation (WCA) and (b) the non-perturbative theory (NP) truncated at level $N=6$ and using $K=6$ Matsubara terms. Parameters used are $T_c=1$, $\Gamma_L=1$, $\Gamma_R=0.025$, $\lambda=0.015$, $\Omega=50$.}
\label{heom-weak-coupling-F2-energy-bias}
\end{figure*}

Using the projectors $\mathcal{P} = |0\rangle\rangle\langle\langle\tilde{0}|$ and $\mathcal{Q} = I - \mathcal{P}$ the steady state can be written
\begin{equation} \label{projected-steady-state}
|0(\chi)\rangle\rangle =  |0\rangle\rangle\ + \mathcal{Q}|0(\chi)\rangle\rangle.
\end{equation}
Additionally, using $\mathcal{H} = \mathcal{Q}\mathcal{H}\mathcal{Q}$ the eigenvalue equation becomes
\begin{equation}
\mathcal{Q}\mathcal{L}(\chi)\mathcal{Q} |0(\chi)\rangle\rangle = [ \lambda_0(\chi) -\Delta\mathcal{H}(\chi) ] |0(\chi)\rangle\rangle.
\end{equation}
Introducing the pseudo-inverse $\mathcal{R} = \mathcal{Q}\mathcal{H}^{-1}\mathcal{Q}$ (more precisely this is the Drazin group generalized inverse \cite{Baiesi2009}) we obtain
\begin{equation}
\mathcal{Q} |0(\chi)\rangle\rangle = \mathcal{R} [ \lambda_0(\chi) - \Delta\mathcal{H}(\chi) ] |0(\chi)\rangle\rangle,
\end{equation}
which on substituting into Eq. \eqref{projected-steady-state} gives us
\begin{equation} \label{steady-state}
|0(\chi)\rangle\rangle = |0\rangle\rangle + \mathcal{R} [ \lambda_0(\chi) - \Delta\mathcal{H}(\chi) ] |0(\chi)\rangle\rangle.
\end{equation}
Expanding $\lambda_0(\chi)$, $|0(\chi)\rangle\rangle$ and $\Delta\mathcal{H}(\chi)$ about $\chi=0$ gives
\begin{eqnarray}
\lambda_0(\chi) &=& \sum^\infty_{n=1} \frac{(i\chi)^n}{n!} \langle\langle I^n \rangle\rangle \\
|0(\chi)\rangle\rangle &=& \sum^\infty_{n=0} \frac{(i\chi)^n}{n!} |0^{(n)}\rangle\rangle \\
\Delta\mathcal{H}(\chi) &=& \sum^\infty_{n=1} \frac{(i\chi)^n}{n!} \mathcal{H}^{(n)}
\end{eqnarray}
From this it is clear that the zero eigenvalue is equivalent to the cumulant generating function of the statistics. Substituting these expansions into Eqs. \eqref{eigenvalue} and \eqref{steady-state} allows calculation of the current cumulants to arbitrary order, though here we only make use of the first and second cumulants:
\begin{equation} \label{current-mean}
 \langle\langle I^1\rangle\rangle = \langle\langle \tilde{0}| \mathcal{H}^{(1)} |0\rangle\rangle,
 \end{equation}
 \begin{equation} \label{current-noise}
 \langle\langle I^2\rangle\rangle = \langle\langle \tilde{0}| \mathcal{H}^{(2)} - 2\mathcal{H}^{(1)}\mathcal{R}\mathcal{H}^{(1)} |0\rangle\rangle
\end{equation}
where the operators $\mathcal{H}^{(n)} = \mathcal{H}_J$.



\begin{figure*}[!t]
\centering
\includegraphics[scale=0.45,trim={1.5cm 0.5cm 1.5cm 1cm}]{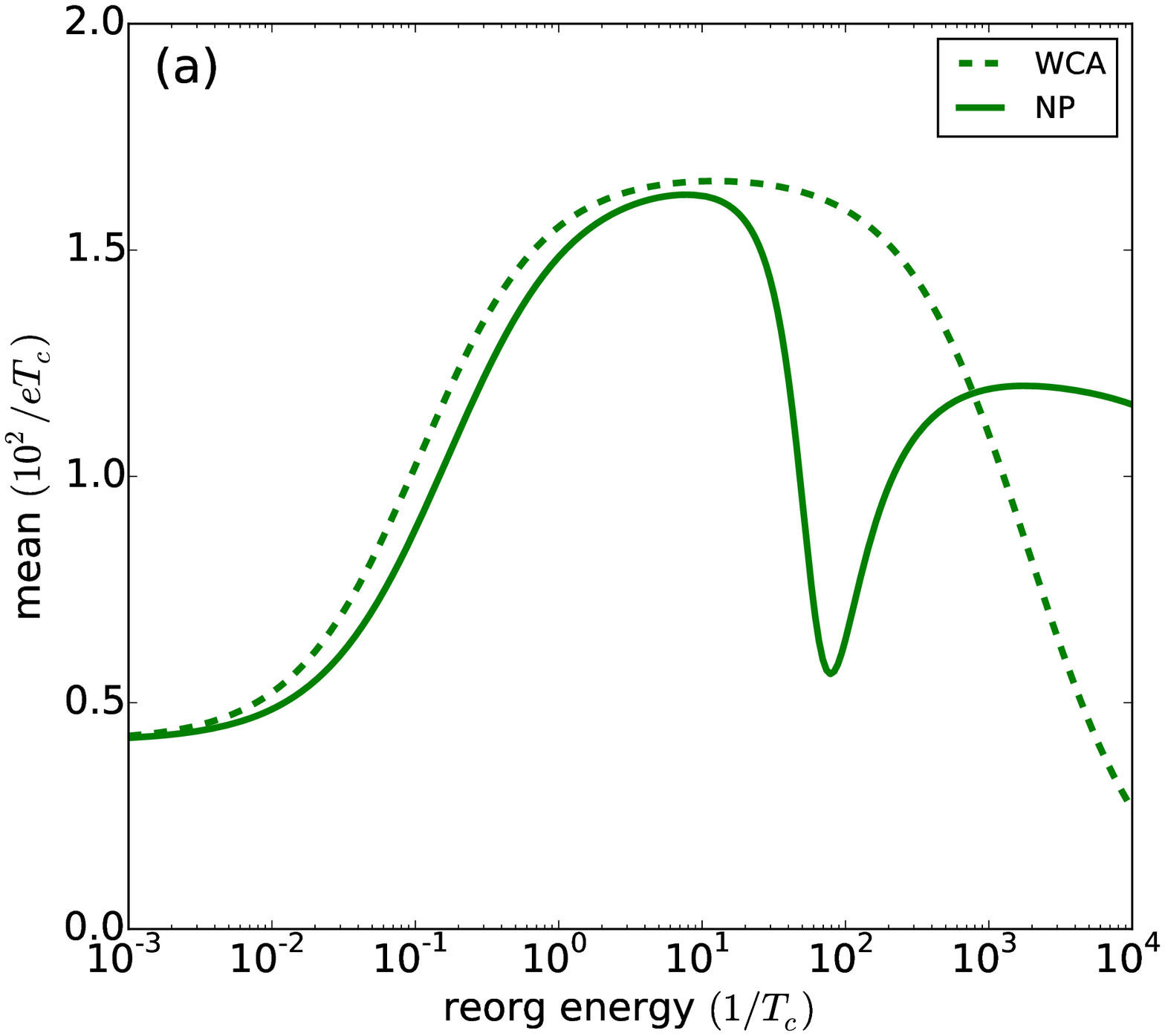}
\hspace{0.5cm}
\includegraphics[scale=0.45,trim={1.5cm 0.5cm 1.5cm -0.4cm}]{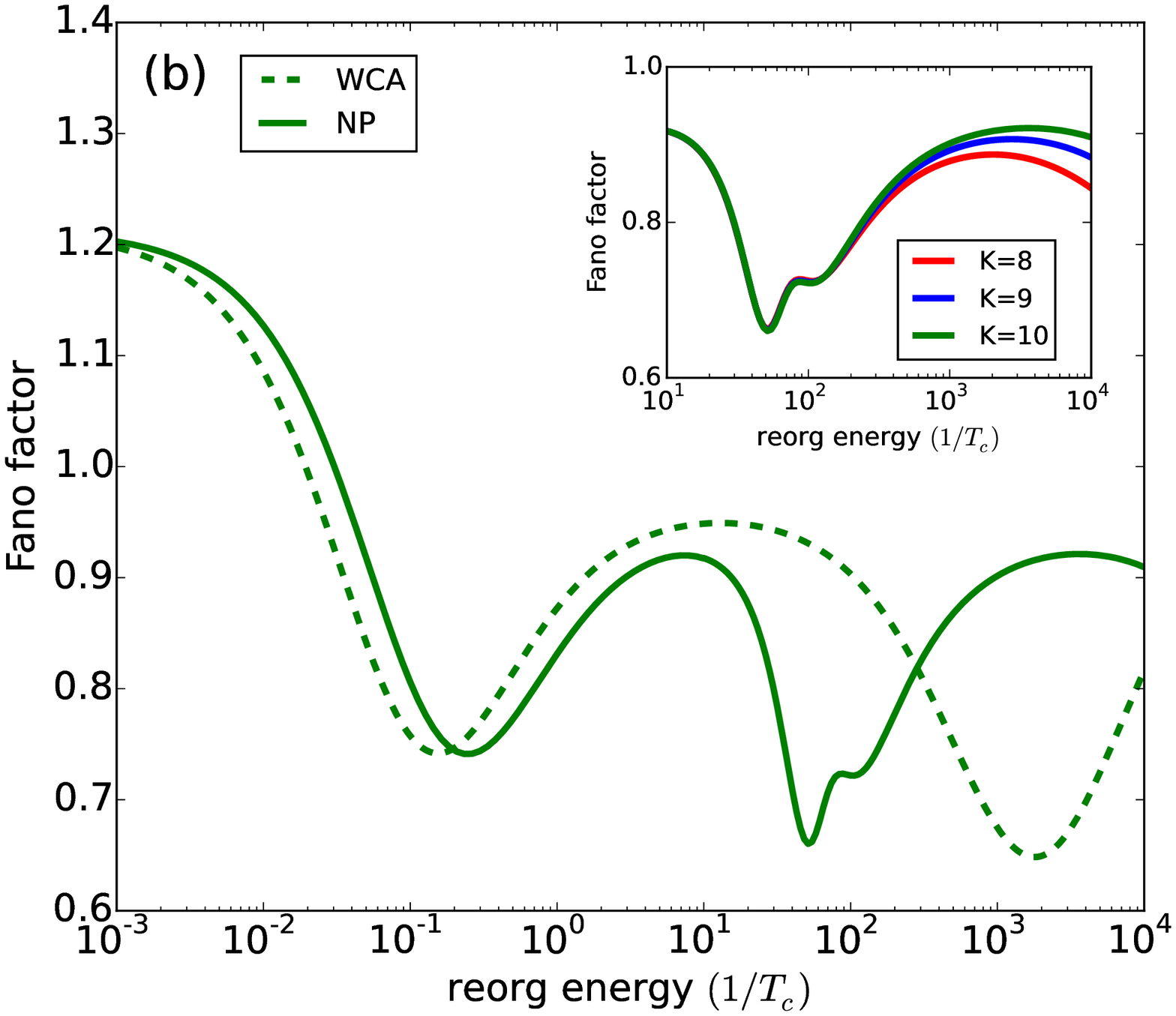}
\caption{(a) Mean current and (b) Fano factor as a function of reorganisation energy. Solid lines are calculated using the non-perturbative (NP) theory truncating the hierarchy at $N=6$ and including $K=10$ Matsubara terms. Dashed lines are calculated using the weak coupling approximation (WCA) with respect to the environment. The inset in (b) shows the convergence of the Fano factor with number of Matsubara terms. Parameters used are $\beta=0.4$, $\epsilon=2$, $T_c=1$, $\Gamma_L=1$, $\Gamma_R=0.025$, $\Omega=50$.}
\label{heom-weak-coupling-F2-reorg-energy}
\end{figure*}

To demonstrate the application of our non-perturbative framework we compare the electron counting statistics to a weak coupling approximation for the system-bath coupling \cite{Brandes2005,Kießlich2007}. The time generator for the perturbative approach is presented in Ref. \cite{Kießlich2007} and reads
\begin{equation}
\mathcal{L} = \begin{pmatrix}
-\Gamma_L & 0 & \Gamma_R e^{i\chi} & 0 & 0 \\
\Gamma_L & 0 & 0 & 0 & 2T_c \\
0 & 0 & -\Gamma_R & 0 & -2T_c \\
0 & \gamma_+ & -\gamma_- & -\frac{\Gamma_R}{2} - \gamma & -\epsilon \\
0 & -T_c & T_c & \epsilon & -\frac{\Gamma_R}{2} - \gamma \\
\end{pmatrix}
\end{equation}
where
\begin{eqnarray}
\gamma =& \frac{2\pi T_c^2}{\Delta^2} J(\Delta) \coth(\frac{\beta\Delta}{2}),  \\
\gamma_{\pm} =& -\frac{\pi\epsilon T_c}{2\Delta^2} J(\Delta) \coth(\frac{\beta\Delta}{2}) \mp \frac{\pi T_c}{2\Delta} J(\Delta)
\end{eqnarray}
are the general expressions for the dephasing rates with an arbitrary spectral density $J(\Delta)$ evaluated at the energy splitting between electronic eigenstates $\Delta = \sqrt{\epsilon^2 + 4T_c^2}$. Rates $\Gamma_L$ and $\Gamma_R$ represent the coupling to the left (source) and right (drain) leads while $\epsilon$ is the energy bias between left and right molecules and $T_c$ is the tunnelling coupling between the molecules. Initially we use a Drude-Lorentz spectral density $J(\omega) = \frac{2\lambda\Omega\omega}{\omega^2 + \Omega^2}$ where $\lambda$ is the reorganisation energy of the bath quantifying the system-bath coupling strength, and $\Omega$ is the cutoff frequency quantifying the bath relaxation time. This form of the spectral density has been successfully used in combination with the hierarchical equations of motion to describe electron transfer in organic molecules \cite{Tanaka2009a,Wang2010}. The central quantity we calculate for electron transport is the Fano factor $F = \frac{\langle\langle I^2\rangle\rangle}{\langle\langle I^1\rangle\rangle}$, which captures the deviation of the current fluctuations from Poissonian statistics.

\begin{figure*}[!th]
\centering
\includegraphics[scale=0.45,trim={1.5cm 0.5cm 1.5cm 1cm}]{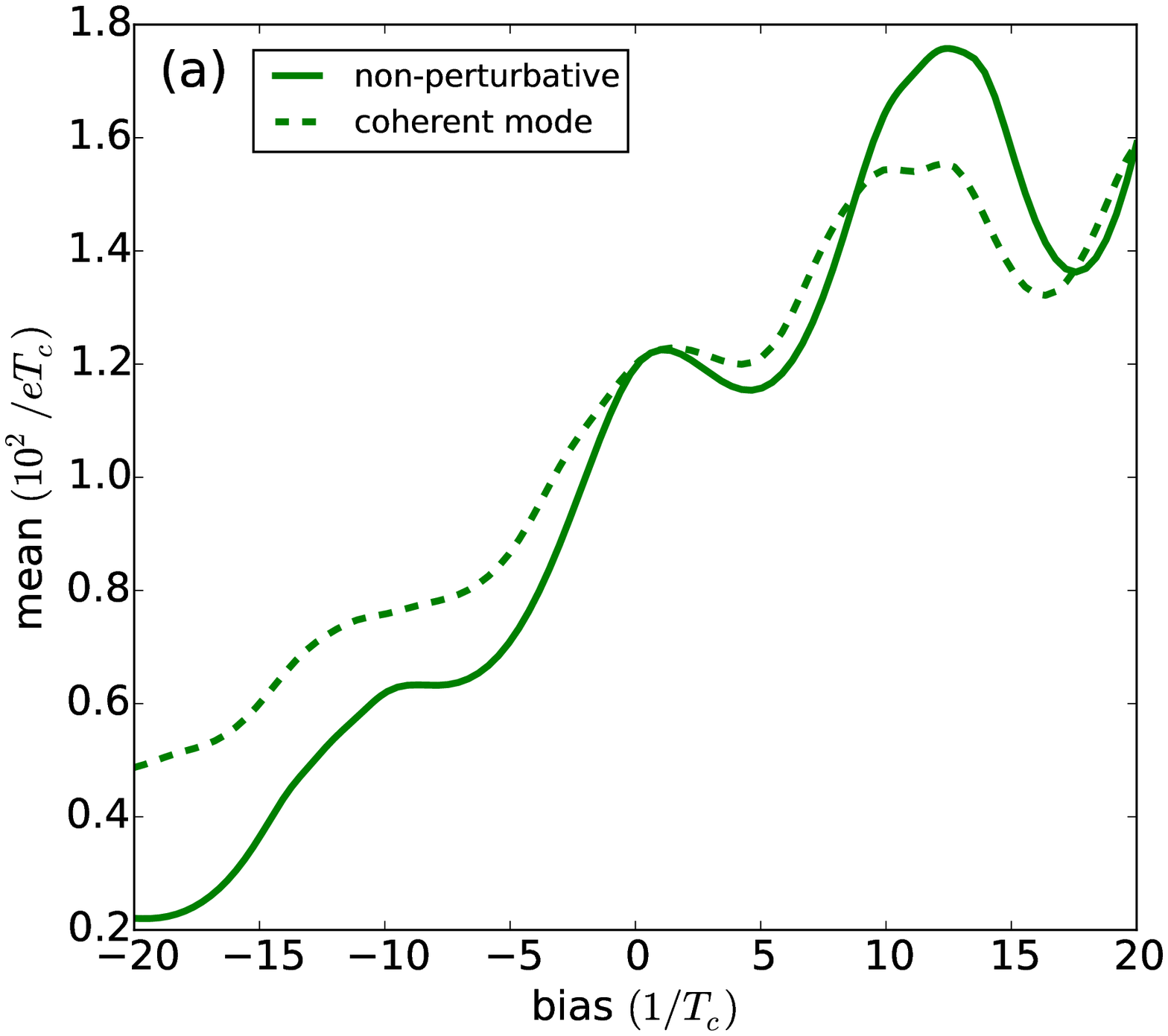}
\hspace{1cm}
\includegraphics[scale=0.45,trim={1.5cm 0.5cm 1.5cm 1cm}]{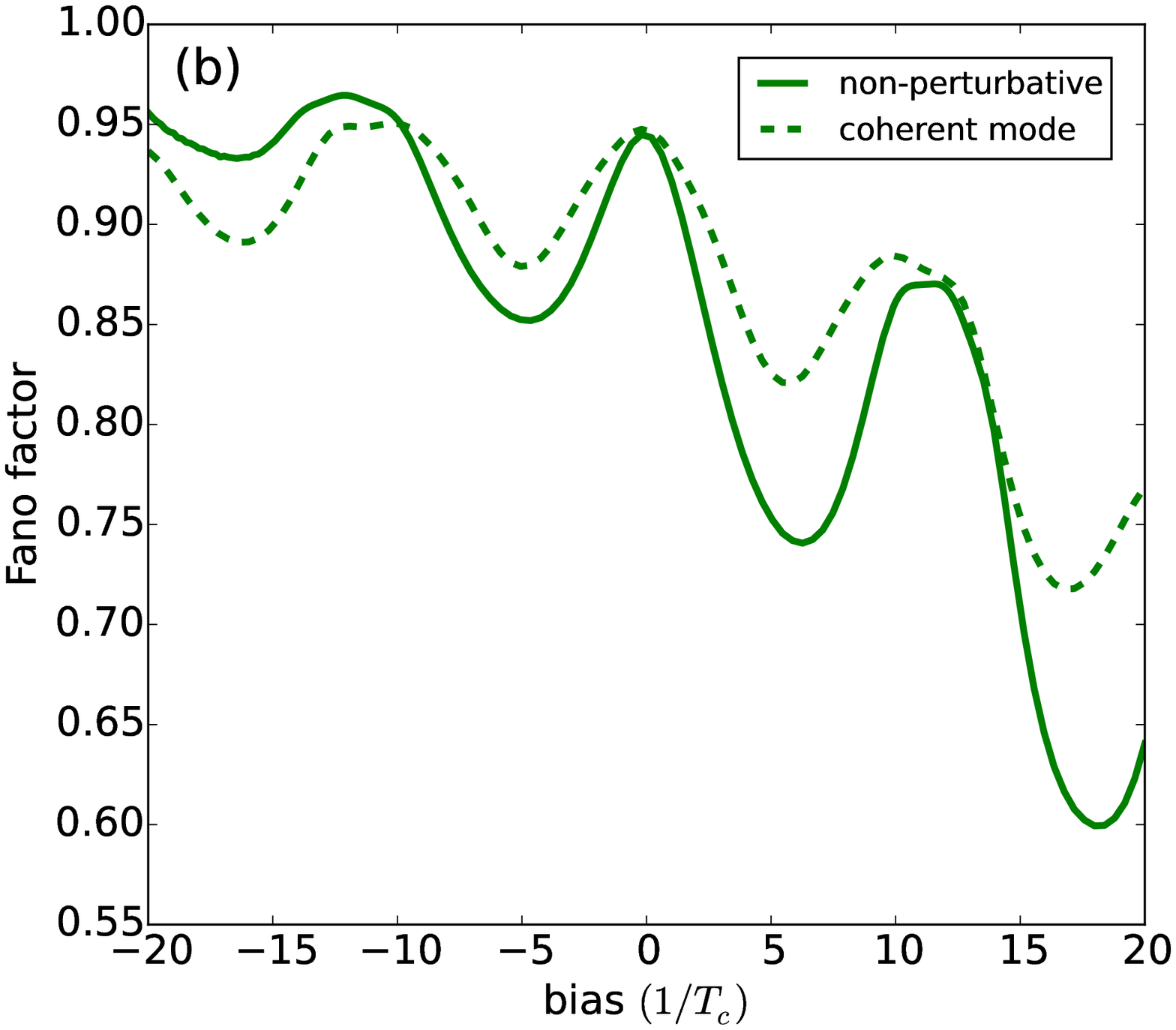}
\caption{(a) Mean current and (b) Fano factor as a function of electronic energy bias comparing an underdamped Brownian oscillator spectral density with a coherently coupled mode model. Solid lines are calculated using the non-perturbative theory while dashed lines are calculated using a Markovian theory including coherent modes in the system Hilbert space. Parameters used are $\beta=0.1$, $T_c=1$, $\Gamma_L=1$, $\Gamma_R=0.025$, $\omega_0=\omega_L=\omega_R=10$, $S=S_L=S_R=0.5$, $\gamma=\gamma_L=\gamma_R=0.5$.}
\label{heom-coherent-mode-F2-bias-UBO}
\end{figure*}

The Fano factor as a function of dimer energy bias $\epsilon$ is shown in Fig. \ref{heom-weak-coupling-F2-energy-bias}. For a double quantum dot system in the absence of a phonon bath the Fano factor is related to the steady state density matrix elements by the expression
\begin{widetext}
\begin{eqnarray}
F^{(2)}(0) = 1 + \frac{2}{\frac{1}{4}\Gamma_R^2\Gamma_L + 2T_c^2\Gamma_L + 4\epsilon^2\Gamma_L + T_c^2\Gamma_R} &\Big[& -T_c^2\Gamma_R\rho_{LL} - \big(\frac{1}{4}\Gamma_R^2\Gamma_L + T_c^2\Gamma_R + 4\epsilon^2\Gamma_L\big)\rho_{RR} \nonumber\\
&+& 4T_c\epsilon\Gamma_L \mbox{Re}(\rho_{LR}) + T_c\Gamma_R\Gamma_L \mbox{Im}(\rho_{LR}) \Big].
\end{eqnarray}
\end{widetext}
It is clear that the Fano factor is sensitive to both steady state populations and coherences with the contribution of the coherences dropping off as the energy bias becomes large. For small couplings to the drain lead ($\Gamma_R \ll 1$) this expression gives super-Poissonian current statistics ($F>1$) indicating charge transfer behaviour similar to the dynamical channel blockade \cite{Kießlich2007}. In this regime delocalization of the charge across the dimer sets up two effective transport channels through either the symmetric or antisymmetric states of the dimer. A difference in the rate of transfer through each channel, in combination with the Coulomb blockade, leads to disordered transport and an increase in current fluctuations. This regime is sensitive to the effects of dephasing and so provides a good model in which to compare the effects of a phonon bath in both perturbative and non-perturbative theories. On introduction of the phonon bath, an asymmetry in the Fano factor with respect to positive or negative energy bias is obtained due to the increased rate of phonon emission relative to phonon absorption. However even for a small reorganisation energy ($\lambda \ll T_c$) there are clear differences between the predictions for the same set of parameters. In the weak system-bath coupling approximation, the interaction with the phonon bath induces dephasing processes which reduce the steady state coherence and consequently the Fano factor. Quantum dynamics in the non-perturbative regime is non-Markovian, where system-bath correlations help sustain electronic coherence for longer time scales. Consequently, a larger non-equilibrium steady state coherence is maintained, leading to larger Fano factors. The influence of this bath dynamics is most marked at higher temperatures ($\beta = 0.1$),  where a qualitative difference in the statistics for negative biases is observed. These results show that even in a regime where the weak-coupling approximation is expected to be valid to describe dynamics, the qualitative features of the zero-noise fluctuations may not be well-captured by these approximations.

The mean current and Fano factor as a function of reorganisation energy for low temperature ($\beta = 0.4$) is shown in Fig. \ref{heom-weak-coupling-F2-reorg-energy}. For small reorganisations energies, where the bath has a weak influence on the electronic system, the weak coupling approximation and the non-perturbative theory give consistent results in both the first two current cumulants. As the reorganisation energy becomes the largest energy scale in the system, qualitative differences start to emerge in the predictions of both theories which signals the breakdown of the weak coupling approximation. This is due to the increased importance of system-bath correlations and the accompanying change in the nature of the electron-phonon relaxation process. The behaviour of both the mean and Fano factor is therefore more counter-intuitive, with the mean in particular exhibiting two maxima rather than the single maxima in the weak coupling approximation. This indicates the presence of two different optimization processes for large and small reorganisation energies with more moderate values leading to relaxation in a renormalized electronic basis such that population localization gives a reduced current output. The inset of Fig. \ref{heom-weak-coupling-F2-reorg-energy} (b) demonstrates the convergence of the Fano factor with increasing Matsubara terms. As coupling to the bath increases more levels of the hierarchy and more Matsubara terms are required for convergence. At a certain reorganisation energy the computational effort will eventually become too great to achieve accurate results.

Coupling to discrete vibrational modes is also an important feature of electron transfer in molecular systems. Hence we investigate the effect of an underdamped Brownian oscillator spectral density on the current and Fano factor at high temperatures in Fig. \ref{heom-coherent-mode-F2-bias-UBO}. This spectral density has the form $J_{BO}(\omega) = \Theta(\omega) \frac{2\Lambda\gamma\omega_0^2 \omega}{(\omega^2 - \omega_0^2)^2 + \gamma^2\omega^2}$ where $\omega_0$ is the frequency of the mode, $\gamma$ is the damping constant and $\Lambda = S\omega_0$ is the reorganisation energy associated with the mode and $S$ is the Huang-Rhys factor. With this spectral density the hierarchical equations of motion take a slightly different form to that presented above \cite{Tanimura2012}. Rather than comparing the non-perturbative theory to the weak coupling approximation, a model in which a damped mode is coherently coupled to each site is used. In this coherent mode model the system Hamiltonian is
\begin{eqnarray}
H_S = &\frac{\epsilon}{2}& (|L\rangle\langle L|-|R\rangle\langle R|) +  T_c(|L\rangle\langle R| + |R\rangle\langle L|) \nonumber \\
&+& \sum_{j=L,R}\hbar\omega_j b^\dagger_j b_j + \sum_{j=L,R} g_j |j\rangle\langle j| (b^\dagger_j + b_j)
\end{eqnarray}
where each mode is coupled to an individual site such that there is no direct interaction with the mode on the adjacent site. The coupling between the electronic and vibrational systems is given by $g_j = \sqrt{S_j}\omega_j$. Damping of the mode is provided using a Lindblad dissipator on the form
\begin{eqnarray}
\mathcal{D}(\rho(t)) = &\sum_{j=L,R}& [ \gamma_j(n(\omega_j)+1) (b_j \rho(t) b_j^\dagger - \frac{1}{2}\{ b_j^\dagger b_j, \rho(t) \}) \nonumber \\
&+& \gamma_j n(\omega_j) (b_j^\dagger \rho(t) b_j - \frac{1}{2}\{ b_j b_j^\dagger, \rho(t) \}) ]
\end{eqnarray}
where the quantities $\gamma_j$ are temperature independent damping rates and $n(\omega_j) = (e^{\beta\omega_j} - 1)^{-1}$ gives the occupation number of mode $j$ at inverse temperature $\beta$. Coupling with the leads is included in the same way in both models. The current and Fano factor are obtained using Eqs. 34 from Ref. \cite{Flindt2010} (the Markovian analogues of Eqs. \eqref{current-mean} and \eqref{current-noise} in this paper) with the same bath parameters used for both frameworks: $\omega_0 = \omega_L = \omega_R$, $S = S_L = S_R$ and $\gamma = \gamma_L = \gamma_R$. Comparing the two models the current and Fano factor have the same qualitative features, with a peak at zero bias and enhancements in both quantities at resonance with the mode frequency. Asymmetry around zero bias is also present due to differences between phonon emission and absorption processes. The main qualitative difference however appears in the position of the peaks around the mode resonances. In the non-perturbative theory the peaks are shifted towards larger energies due to renormalisation of the electronic energy levels which is not fully captured in the coherent mode model.

A limitation to the non-perturbative and coherent mode models presented here lies in the validity of the description of the coupling between the electronic system and the leads. 
In the presence of strong electron-phonon interaction the leads may not interact with the electronic system independently of the phonon bath, rather with vibronic states. While this assumption has been used previously for counting statistics calculations in similar systems with arbitrary electron-phonon coupling \cite{Braggio2009,Santamore2013}, further developments to the non-perturbative theory will seek to investigate the extent of this issue.

In summary, we have have put forward a non-perturbative method to calculate electron counting statistics using the hierarchical equations of motion. Single molecule or single protein junctions in conjunction with single-electron counting devices \cite{Nishiguchi2009} will allow investigation of the statistics of the current passing through organic molecules thereby providing a powerful tool to complement transient spectroscopic studies. This in turn will lead to a deeper understanding of the microscopic mechanisms underlying electron transfer in a variety of organic systems while at the same time giving valuable insight for the design of reliable and better performing single-molecule devices.

\section*{Acknowledgements}
The authors would like to thank Clive Emary for useful discussions. Financial support from the Engineering and Physical Sciences Research Council is gratefully acknowledged.


\begin{thebibliography}{32}%
\makeatletter
\providecommand \@ifxundefined [1]{%
 \@ifx{#1\undefined}
}%
\providecommand \@ifnum [1]{%
 \ifnum #1\expandafter \@firstoftwo
 \else \expandafter \@secondoftwo
 \fi
}%
\providecommand \@ifx [1]{%
 \ifx #1\expandafter \@firstoftwo
 \else \expandafter \@secondoftwo
 \fi
}%
\providecommand \natexlab [1]{#1}%
\providecommand \enquote  [1]{``#1''}%
\providecommand \bibnamefont  [1]{#1}%
\providecommand \bibfnamefont [1]{#1}%
\providecommand \citenamefont [1]{#1}%
\providecommand \href@noop [0]{\@secondoftwo}%
\providecommand \href [0]{\begingroup \@sanitize@url \@href}%
\providecommand \@href[1]{\@@startlink{#1}\@@href}%
\providecommand \@@href[1]{\endgroup#1\@@endlink}%
\providecommand \@sanitize@url [0]{\catcode `\\12\catcode `\$12\catcode
  `\&12\catcode `\#12\catcode `\^12\catcode `\_12\catcode `\%12\relax}%
\providecommand \@@startlink[1]{}%
\providecommand \@@endlink[0]{}%
\providecommand \url  [0]{\begingroup\@sanitize@url \@url }%
\providecommand \@url [1]{\endgroup\@href {#1}{\urlprefix }}%
\providecommand \urlprefix  [0]{URL }%
\providecommand \Eprint [0]{\href }%
\providecommand \doibase [0]{http://dx.doi.org/}%
\providecommand \selectlanguage [0]{\@gobble}%
\providecommand \bibinfo  [0]{\@secondoftwo}%
\providecommand \bibfield  [0]{\@secondoftwo}%
\providecommand \translation [1]{[#1]}%
\providecommand \BibitemOpen [0]{}%
\providecommand \bibitemStop [0]{}%
\providecommand \bibitemNoStop [0]{.\EOS\space}%
\providecommand \EOS [0]{\spacefactor3000\relax}%
\providecommand \BibitemShut  [1]{\csname bibitem#1\endcsname}%
\let\auto@bib@innerbib\@empty
\bibitem [{\citenamefont {Gray}\ and\ \citenamefont
  {Winkler}(1996)}]{Gray1996}%
  \BibitemOpen
  \bibfield  {author} {\bibinfo {author} {\bibfnamefont {H.~B.}\ \bibnamefont
  {Gray}}\ and\ \bibinfo {author} {\bibfnamefont {J.~R.}\ \bibnamefont
  {Winkler}},\ }\href {\doibase 10.1146/annurev.bi.65.070196.002541} {\bibfield
   {journal} {\bibinfo  {journal} {Annu. Rev. Biochem.}\ }\textbf {\bibinfo
  {volume} {837}},\ \bibinfo {pages} {1973} (\bibinfo {year}
  {1996})}\BibitemShut {NoStop}%
\bibitem [{\citenamefont {Fink}\ and\ \citenamefont
  {Sch\"{o}nenberger}(1999)}]{Fink1999}%
  \BibitemOpen
  \bibfield  {author} {\bibinfo {author} {\bibfnamefont {H.-W.}\ \bibnamefont
  {Fink}}\ and\ \bibinfo {author} {\bibfnamefont {C.}~\bibnamefont
  {Sch\"{o}nenberger}},\ }\href@noop {} {\bibfield  {journal} {\bibinfo
  {journal} {Nature}\ }\textbf {\bibinfo {volume} {398}},\ \bibinfo {pages}
  {407} (\bibinfo {year} {1999})}\BibitemShut {NoStop}%
\bibitem [{\citenamefont {Gerster}\ \emph {et~al.}(2012)\citenamefont
  {Gerster}, \citenamefont {Reichert}, \citenamefont {Bi}, \citenamefont
  {Barth}, \citenamefont {Kaniber}, \citenamefont {Holleitner}, \citenamefont
  {Visoly-Fisher}, \citenamefont {Sergani},\ and\ \citenamefont
  {Carmeli}}]{Gerster2012}%
  \BibitemOpen
  \bibfield  {author} {\bibinfo {author} {\bibfnamefont {D.}~\bibnamefont
  {Gerster}}, \bibinfo {author} {\bibfnamefont {J.}~\bibnamefont {Reichert}},
  \bibinfo {author} {\bibfnamefont {H.}~\bibnamefont {Bi}}, \bibinfo {author}
  {\bibfnamefont {J.}~\bibnamefont {Barth}}, \bibinfo {author} {\bibfnamefont
  {S.}~\bibnamefont {Kaniber}}, \bibinfo {author} {\bibfnamefont
  {A.}~\bibnamefont {Holleitner}}, \bibinfo {author} {\bibfnamefont
  {I.}~\bibnamefont {Visoly-Fisher}}, \bibinfo {author} {\bibfnamefont
  {S.}~\bibnamefont {Sergani}}, \ and\ \bibinfo {author} {\bibfnamefont
  {I.}~\bibnamefont {Carmeli}},\ }\href {\doibase 10.1038/nnano.2012.165}
  {\bibfield  {journal} {\bibinfo  {journal} {Nature Nanotech.}\ }\textbf
  {\bibinfo {volume} {7}},\ \bibinfo {pages} {673} (\bibinfo {year}
  {2012})}\BibitemShut {NoStop}%
\bibitem [{\citenamefont {Tao}(2006)}]{Tao2006}%
  \BibitemOpen
  \bibfield  {author} {\bibinfo {author} {\bibfnamefont {N.~J.}\ \bibnamefont
  {Tao}},\ }\href {\doibase 10.1038/nnano.2006.130} {\bibfield  {journal}
  {\bibinfo  {journal} {Nature Nanotech.}\ }\textbf {\bibinfo {volume} {1}},\
  \bibinfo {pages} {173} (\bibinfo {year} {2006})}\BibitemShut {NoStop}%
\bibitem [{\citenamefont {Selzer}\ and\ \citenamefont
  {Allara}(2006)}]{Selzer2006}%
  \BibitemOpen
  \bibfield  {author} {\bibinfo {author} {\bibfnamefont {Y.}~\bibnamefont
  {Selzer}}\ and\ \bibinfo {author} {\bibfnamefont {D.~L.}\ \bibnamefont
  {Allara}},\ }\href {\doibase 10.1146/annurev.physchem.57.032905.104709}
  {\bibfield  {journal} {\bibinfo  {journal} {Ann. Rev. Phys. Chem.}\ }\textbf
  {\bibinfo {volume} {57}},\ \bibinfo {pages} {593} (\bibinfo {year}
  {2006})}\BibitemShut {NoStop}%
\bibitem [{\citenamefont {Levitov}\ and\ \citenamefont
  {Lesovik}(1993)}]{Levitov1993}%
  \BibitemOpen
  \bibfield  {author} {\bibinfo {author} {\bibfnamefont {L.}~\bibnamefont
  {Levitov}}\ and\ \bibinfo {author} {\bibfnamefont {G.}~\bibnamefont
  {Lesovik}},\ }\href
  {http://www.jetpletters.ac.ru/ps/1186/article{\_}17907.pdf} {\bibfield
  {journal} {\bibinfo  {journal} {JETP Letters}\ }\textbf {\bibinfo {volume}
  {58}} (\bibinfo {year} {1993})}\BibitemShut {NoStop}%
\bibitem [{\citenamefont {Blanter}\ and\ \citenamefont
  {B{\"{u}}ttiker}(2000)}]{Blanter2000}%
  \BibitemOpen
  \bibfield  {author} {\bibinfo {author} {\bibfnamefont {Y.}~\bibnamefont
  {Blanter}}\ and\ \bibinfo {author} {\bibfnamefont {M.}~\bibnamefont
  {B{\"{u}}ttiker}},\ }\href
  {http://www.sciencedirect.com/science/article/pii/S0370157399001234}
  {\bibfield  {journal} {\bibinfo  {journal} {Physics Reports}\ }\textbf
  {\bibinfo {volume} {336}},\ \bibinfo {pages} {1} (\bibinfo {year}
  {2000})}\BibitemShut {NoStop}%
\bibitem [{\citenamefont {Marcos}\ \emph {et~al.}(2010)\citenamefont {Marcos},
  \citenamefont {Emary}, \citenamefont {Brandes},\ and\ \citenamefont
  {Aguado}}]{Marcos2010}%
  \BibitemOpen
  \bibfield  {author} {\bibinfo {author} {\bibfnamefont {D.}~\bibnamefont
  {Marcos}}, \bibinfo {author} {\bibfnamefont {C.}~\bibnamefont {Emary}},
  \bibinfo {author} {\bibfnamefont {T.}~\bibnamefont {Brandes}}, \ and\
  \bibinfo {author} {\bibfnamefont {R.}~\bibnamefont {Aguado}},\ }\href@noop {}
  {\bibfield  {journal} {\bibinfo  {journal} {NJP}\ }\textbf {\bibinfo {volume}
  {12}} (\bibinfo {year} {2010})}\BibitemShut {NoStop}%
\bibitem [{\citenamefont {Kie{\ss}lich}\ \emph {et~al.}(2007)\citenamefont
  {Kie{\ss}lich}, \citenamefont {Sch{\"{o}}ll}, \citenamefont {Brandes},
  \citenamefont {Hohls},\ and\ \citenamefont {Haug}}]{Kießlich2007}%
  \BibitemOpen
  \bibfield  {author} {\bibinfo {author} {\bibfnamefont {G.}~\bibnamefont
  {Kie{\ss}lich}}, \bibinfo {author} {\bibfnamefont {E.}~\bibnamefont
  {Sch{\"{o}}ll}}, \bibinfo {author} {\bibfnamefont {T.}~\bibnamefont
  {Brandes}}, \bibinfo {author} {\bibfnamefont {F.}~\bibnamefont {Hohls}}, \
  and\ \bibinfo {author} {\bibfnamefont {R.}~\bibnamefont {Haug}},\ }\href
  {\doibase 10.1103/PhysRevLett.99.206602} {\bibfield  {journal} {\bibinfo
  {journal} {PRL}\ }\textbf {\bibinfo {volume} {99}},\ \bibinfo {pages}
  {206602} (\bibinfo {year} {2007})}\BibitemShut {NoStop}%
\bibitem [{\citenamefont {Belzig}(2005)}]{Belzig2005}%
  \BibitemOpen
  \bibfield  {author} {\bibinfo {author} {\bibfnamefont {W.}~\bibnamefont
  {Belzig}},\ }\href {\doibase 10.1103/PhysRevB.71.161301} {\bibfield
  {journal} {\bibinfo  {journal} {Phys. Rev. B}\ }\textbf {\bibinfo {volume}
  {71}},\ \bibinfo {pages} {1} (\bibinfo {year} {2005})}\BibitemShut {NoStop}%
\bibitem [{\citenamefont {Haupt}\ \emph {et~al.}(2006)\citenamefont {Haupt},
  \citenamefont {Cavaliere}, \citenamefont {Fazio},\ and\ \citenamefont
  {Sassetti}}]{Haupt2006}%
  \BibitemOpen
  \bibfield  {author} {\bibinfo {author} {\bibfnamefont {F.}~\bibnamefont
  {Haupt}}, \bibinfo {author} {\bibfnamefont {F.}~\bibnamefont {Cavaliere}},
  \bibinfo {author} {\bibfnamefont {R.}~\bibnamefont {Fazio}}, \ and\ \bibinfo
  {author} {\bibfnamefont {M.}~\bibnamefont {Sassetti}},\ }\href@noop {}
  {\bibfield  {journal} {\bibinfo  {journal} {Phys. Rev. B}\ }\textbf {\bibinfo
  {volume} {74}},\ \bibinfo {pages} {205328} (\bibinfo {year}
  {2006})}\BibitemShut {NoStop}%
\bibitem [{\citenamefont {Brandes}(2005)}]{Brandes2005}%
  \BibitemOpen
  \bibfield  {author} {\bibinfo {author} {\bibfnamefont {T.}~\bibnamefont
  {Brandes}},\ }\href {\doibase 10.1016/j.physrep.2004.12.002} {\bibfield
  {journal} {\bibinfo  {journal} {Physics Reports}\ }\textbf {\bibinfo {volume}
  {408}},\ \bibinfo {pages} {315} (\bibinfo {year} {2005})}\BibitemShut
  {NoStop}%
\bibitem [{\citenamefont {Braggio}\ \emph
  {et~al.}(2009{\natexlab{a}})\citenamefont {Braggio}, \citenamefont
  {K\"{o}nig},\ and\ \citenamefont {Fazio}}]{Braggio2006}%
  \BibitemOpen
  \bibfield  {author} {\bibinfo {author} {\bibfnamefont {A.}~\bibnamefont
  {Braggio}}, \bibinfo {author} {\bibfnamefont {J.}~\bibnamefont {K\"{o}nig}},
  \ and\ \bibinfo {author} {\bibfnamefont {R.}~\bibnamefont {Fazio}},\
  }\href@noop {} {\bibfield  {journal} {\bibinfo  {journal} {PRL}\ }\textbf
  {\bibinfo {volume} {96}},\ \bibinfo {pages} {026805} (\bibinfo {year}
  {2009}{\natexlab{a}})}\BibitemShut {NoStop}%
\bibitem [{\citenamefont {Flindt}\ \emph {et~al.}(2008)\citenamefont {Flindt},
  \citenamefont {Novotn{\'{y}}}, \citenamefont {Braggio}, \citenamefont
  {Sassetti},\ and\ \citenamefont {Jauho}}]{Flindt2008}%
  \BibitemOpen
  \bibfield  {author} {\bibinfo {author} {\bibfnamefont {C.}~\bibnamefont
  {Flindt}}, \bibinfo {author} {\bibfnamefont {T.}~\bibnamefont
  {Novotn{\'{y}}}}, \bibinfo {author} {\bibfnamefont {A.}~\bibnamefont
  {Braggio}}, \bibinfo {author} {\bibfnamefont {M.}~\bibnamefont {Sassetti}}, \
  and\ \bibinfo {author} {\bibfnamefont {A.-P.}\ \bibnamefont {Jauho}},\ }\href
  {\doibase 10.1103/PhysRevLett.100.150601} {\bibfield  {journal} {\bibinfo
  {journal} {PRL}\ }\textbf {\bibinfo {volume} {100}},\ \bibinfo {pages}
  {150601} (\bibinfo {year} {2008})}\BibitemShut {NoStop}%
\bibitem [{\citenamefont {Flindt}\ \emph {et~al.}(2010)\citenamefont {Flindt},
  \citenamefont {Novotn{\'{y}}}, \citenamefont {Braggio},\ and\ \citenamefont
  {Jauho}}]{Flindt2010}%
  \BibitemOpen
  \bibfield  {author} {\bibinfo {author} {\bibfnamefont {C.}~\bibnamefont
  {Flindt}}, \bibinfo {author} {\bibfnamefont {T.}~\bibnamefont
  {Novotn{\'{y}}}}, \bibinfo {author} {\bibfnamefont {A.}~\bibnamefont
  {Braggio}}, \ and\ \bibinfo {author} {\bibfnamefont {A.-P.}\ \bibnamefont
  {Jauho}},\ }\href {\doibase 10.1103/PhysRevB.82.155407} {\bibfield  {journal}
  {\bibinfo  {journal} {Phys. Rev. B}\ }\textbf {\bibinfo {volume} {82}},\
  \bibinfo {pages} {155407} (\bibinfo {year} {2010})}\BibitemShut {NoStop}%
\bibitem [{\citenamefont {Braggio}\ \emph
  {et~al.}(2009{\natexlab{b}})\citenamefont {Braggio}, \citenamefont {Flindt},\
  and\ \citenamefont {Novotn{\'{y}}}}]{Braggio2009}%
  \BibitemOpen
  \bibfield  {author} {\bibinfo {author} {\bibfnamefont {A.}~\bibnamefont
  {Braggio}}, \bibinfo {author} {\bibfnamefont {C.}~\bibnamefont {Flindt}}, \
  and\ \bibinfo {author} {\bibfnamefont {T.}~\bibnamefont {Novotn{\'{y}}}},\
  }\href@noop {} {\bibfield  {journal} {\bibinfo  {journal} {J. Stat. Mech.}\
  }\textbf {\bibinfo {volume} {P01048}} (\bibinfo {year}
  {2009}{\natexlab{b}})}\BibitemShut {NoStop}%
\bibitem [{\citenamefont {Cerillo}\ \emph {et~al.}(2016)\citenamefont
  {Cerillo}, \citenamefont {Buser},\ and\ \citenamefont
  {Brandes}}]{Cerillo2016}%
  \BibitemOpen
  \bibfield  {author} {\bibinfo {author} {\bibfnamefont {J.}~\bibnamefont
  {Cerillo}}, \bibinfo {author} {\bibfnamefont {M.}~\bibnamefont {Buser}}, \
  and\ \bibinfo {author} {\bibfnamefont {T.}~\bibnamefont {Brandes}},\
  }\href@noop {} {\bibfield  {journal} {\bibinfo  {journal} {Phys. Rev. B}\
  }\textbf {\bibinfo {volume} {94}},\ \bibinfo {pages} {214308} (\bibinfo
  {year} {2016})}\BibitemShut {NoStop}%
\bibitem [{\citenamefont {Tanimura}(2006)}]{Tanimura2006}%
  \BibitemOpen
  \bibfield  {author} {\bibinfo {author} {\bibfnamefont {Y.}~\bibnamefont
  {Tanimura}},\ }\href {http://journals.jps.jp/doi/abs/10.1143/JPSJ.75.082001}
  {\bibfield  {journal} {\bibinfo  {journal} {J. Phys. Soc. Jap.}\ }\textbf
  {\bibinfo {volume} {75}},\ \bibinfo {pages} {082001} (\bibinfo {year}
  {2006})}\BibitemShut {NoStop}%
\bibitem [{\citenamefont {Ishizaki}\ and\ \citenamefont
  {Fleming}(2009)}]{Ishizaki2009}%
  \BibitemOpen
  \bibfield  {author} {\bibinfo {author} {\bibfnamefont {A.}~\bibnamefont
  {Ishizaki}}\ and\ \bibinfo {author} {\bibfnamefont {G.~R.}\ \bibnamefont
  {Fleming}},\ }\href {\doibase 10.1063/1.3155372} {\bibfield  {journal}
  {\bibinfo  {journal} {J. Chem. Phys.}\ }\textbf {\bibinfo {volume} {130}},\
  \bibinfo {pages} {234111} (\bibinfo {year} {2009})}\BibitemShut {NoStop}%
\bibitem [{\citenamefont {Tanimura}(2012)}]{Tanimura2012}%
  \BibitemOpen
  \bibfield  {author} {\bibinfo {author} {\bibfnamefont {Y.}~\bibnamefont
  {Tanimura}},\ }\href@noop {} {\bibfield  {journal} {\bibinfo  {journal} {J.
  Chem. Phys.}\ }\textbf {\bibinfo {volume} {137}} (\bibinfo {year}
  {2012})}\BibitemShut {NoStop}%
\bibitem [{\citenamefont {Kreisbeck}\ \emph {et~al.}(2011)\citenamefont
  {Kreisbeck}, \citenamefont {Kramer}, \citenamefont {Rodr{\'{i}}guez},\ and\
  \citenamefont {Hein}}]{Kreisbeck2011}%
  \BibitemOpen
  \bibfield  {author} {\bibinfo {author} {\bibfnamefont {C.}~\bibnamefont
  {Kreisbeck}}, \bibinfo {author} {\bibfnamefont {T.}~\bibnamefont {Kramer}},
  \bibinfo {author} {\bibfnamefont {M.}~\bibnamefont {Rodr{\'{i}}guez}}, \ and\
  \bibinfo {author} {\bibfnamefont {B.}~\bibnamefont {Hein}},\ }\href {\doibase
  10.1021/ct200126d} {\bibfield  {journal} {\bibinfo  {journal} {J. Chem. Theo.
  Comp.}\ }\textbf {\bibinfo {volume} {7}},\ \bibinfo {pages} {2166} (\bibinfo
  {year} {2011})}\BibitemShut {NoStop}%
\bibitem [{\citenamefont {Flindt}\ \emph {et~al.}(2005)\citenamefont {Flindt},
  \citenamefont {Novotn{\'{y}}},\ and\ \citenamefont
  {a.~P~Jauho}}]{Flindt2005}%
  \BibitemOpen
  \bibfield  {author} {\bibinfo {author} {\bibfnamefont {C.}~\bibnamefont
  {Flindt}}, \bibinfo {author} {\bibfnamefont {T.}~\bibnamefont
  {Novotn{\'{y}}}}, \ and\ \bibinfo {author} {\bibnamefont {a.~P~Jauho}},\
  }\href {\doibase 10.1209/epl/i2004-10351-x} {\bibfield  {journal} {\bibinfo
  {journal} {Europhys. Lett.}\ }\textbf {\bibinfo {volume} {69}},\ \bibinfo
  {pages} {475} (\bibinfo {year} {2005})}\BibitemShut {NoStop}%
\bibitem [{\citenamefont {Harbola}\ \emph {et~al.}(2006)\citenamefont
  {Harbola}, \citenamefont {Esposito},\ and\ \citenamefont
  {Mukamel}}]{Harbola2006}%
  \BibitemOpen
  \bibfield  {author} {\bibinfo {author} {\bibfnamefont {U.}~\bibnamefont
  {Harbola}}, \bibinfo {author} {\bibfnamefont {M.}~\bibnamefont {Esposito}}, \
  and\ \bibinfo {author} {\bibfnamefont {S.}~\bibnamefont {Mukamel}},\ }\href
  {\doibase 10.1103/PhysRevB.74.235309} {\bibfield  {journal} {\bibinfo
  {journal} {Phys. Rev. B}\ }\textbf {\bibinfo {volume} {74}},\ \bibinfo
  {pages} {235309} (\bibinfo {year} {2006})}\BibitemShut {NoStop}%
\bibitem [{\citenamefont {Zhu}\ \emph {et~al.}(2012)\citenamefont {Zhu},
  \citenamefont {Liu}, \citenamefont {Xie},\ and\ \citenamefont
  {Shi}}]{Zhu2012}%
  \BibitemOpen
  \bibfield  {author} {\bibinfo {author} {\bibfnamefont {L.}~\bibnamefont
  {Zhu}}, \bibinfo {author} {\bibfnamefont {H.}~\bibnamefont {Liu}}, \bibinfo
  {author} {\bibfnamefont {W.}~\bibnamefont {Xie}}, \ and\ \bibinfo {author}
  {\bibfnamefont {Q.}~\bibnamefont {Shi}},\ }\href@noop {} {\bibfield
  {journal} {\bibinfo  {journal} {J. Chem. Phys.}\ }\textbf {\bibinfo {volume}
  {137}},\ \bibinfo {pages} {194106} (\bibinfo {year} {2012})}\BibitemShut
  {NoStop}%
\bibitem [{\citenamefont {Shi}\ \emph {et~al.}(2009)\citenamefont {Shi},
  \citenamefont {Chen}, \citenamefont {Nan}, \citenamefont {Xu}, \citenamefont
  {Yan}, \citenamefont {Shi}, \citenamefont {Chen}, \citenamefont {Nan},
  \citenamefont {Xu},\ and\ \citenamefont {Yan}}]{Shi2009}%
  \BibitemOpen
  \bibfield  {author} {\bibinfo {author} {\bibfnamefont {Q.}~\bibnamefont
  {Shi}}, \bibinfo {author} {\bibfnamefont {L.}~\bibnamefont {Chen}}, \bibinfo
  {author} {\bibfnamefont {G.}~\bibnamefont {Nan}}, \bibinfo {author}
  {\bibfnamefont {R.-x.}\ \bibnamefont {Xu}}, \bibinfo {author} {\bibfnamefont
  {Y.}~\bibnamefont {Yan}}, \bibinfo {author} {\bibfnamefont {Q.}~\bibnamefont
  {Shi}}, \bibinfo {author} {\bibfnamefont {L.}~\bibnamefont {Chen}}, \bibinfo
  {author} {\bibfnamefont {G.}~\bibnamefont {Nan}}, \bibinfo {author}
  {\bibfnamefont {R.-x.}\ \bibnamefont {Xu}}, \ and\ \bibinfo {author}
  {\bibfnamefont {Y.}~\bibnamefont {Yan}},\ }\href@noop {} {\bibfield
  {journal} {\bibinfo  {journal} {J. Chem. Phys.}\ }\textbf {\bibinfo {volume}
  {130}} (\bibinfo {year} {2009})}\BibitemShut {NoStop}%
\bibitem [{\citenamefont {Ishizaki}\ and\ \citenamefont
  {Tanimura}(2005)}]{Ishizaki2005}%
  \BibitemOpen
  \bibfield  {author} {\bibinfo {author} {\bibfnamefont {A.}~\bibnamefont
  {Ishizaki}}\ and\ \bibinfo {author} {\bibfnamefont {Y.}~\bibnamefont
  {Tanimura}},\ }\href@noop {} {\bibfield  {journal} {\bibinfo  {journal} {J.
  Phys. Soc. Jpn.}\ }\textbf {\bibinfo {volume} {74}},\ \bibinfo {pages} {3131}
  (\bibinfo {year} {2005})}\BibitemShut {NoStop}%
\bibitem [{\citenamefont {Bagrets}\ and\ \citenamefont
  {Nazarov}(2003)}]{Bagrets2003}%
  \BibitemOpen
  \bibfield  {author} {\bibinfo {author} {\bibfnamefont {D.}~\bibnamefont
  {Bagrets}}\ and\ \bibinfo {author} {\bibfnamefont {Y.}~\bibnamefont
  {Nazarov}},\ }\href {\doibase 10.1103/PhysRevB.67.085316} {\bibfield
  {journal} {\bibinfo  {journal} {Phys. Rev. B}\ }\textbf {\bibinfo {volume}
  {67}},\ \bibinfo {pages} {085316} (\bibinfo {year} {2003})}\BibitemShut
  {NoStop}%
\bibitem [{\citenamefont {Baiesi}\ \emph {et~al.}(2009)\citenamefont {Baiesi},
  \citenamefont {Maes},\ and\ \citenamefont {Netoˇ}}]{Baiesi2009}%
  \BibitemOpen
  \bibfield  {author} {\bibinfo {author} {\bibfnamefont {M.}~\bibnamefont
  {Baiesi}}, \bibinfo {author} {\bibfnamefont {C.}~\bibnamefont {Maes}}, \ and\
  \bibinfo {author} {\bibfnamefont {K.}~\bibnamefont {Netoˇ}},\ }\href
  {\doibase 10.1007/s10955-009-9723-3} {\bibfield  {journal} {\bibinfo
  {journal} {J. Stat. Phys.}\ }\textbf {\bibinfo {volume} {135}},\ \bibinfo
  {pages} {57} (\bibinfo {year} {2009})}\BibitemShut {NoStop}%
\bibitem [{\citenamefont {Tanaka}\ and\ \citenamefont
  {Tanimura}(2009)}]{Tanaka2009a}%
  \BibitemOpen
  \bibfield  {author} {\bibinfo {author} {\bibfnamefont {M.}~\bibnamefont
  {Tanaka}}\ and\ \bibinfo {author} {\bibfnamefont {Y.}~\bibnamefont
  {Tanimura}},\ }\href@noop {} {\bibfield  {journal} {\bibinfo  {journal} {J.
  Phys. Soc. Jap.}\ }\textbf {\bibinfo {volume} {78}},\ \bibinfo {pages}
  {073802} (\bibinfo {year} {2009})}\BibitemShut {NoStop}%
\bibitem [{\citenamefont {Wang}\ \emph {et~al.}(2010)\citenamefont {Wang},
  \citenamefont {Chen}, \citenamefont {Zheng}, \citenamefont {Wang},\ and\
  \citenamefont {Shi}}]{Wang2010}%
  \BibitemOpen
  \bibfield  {author} {\bibinfo {author} {\bibfnamefont {D.}~\bibnamefont
  {Wang}}, \bibinfo {author} {\bibfnamefont {L.}~\bibnamefont {Chen}}, \bibinfo
  {author} {\bibfnamefont {R.}~\bibnamefont {Zheng}}, \bibinfo {author}
  {\bibfnamefont {L.}~\bibnamefont {Wang}}, \ and\ \bibinfo {author}
  {\bibfnamefont {Q.}~\bibnamefont {Shi}},\ }\href@noop {} {\bibfield
  {journal} {\bibinfo  {journal} {J. Chem. Phys.}\ }\textbf {\bibinfo {volume}
  {132}},\ \bibinfo {pages} {081101} (\bibinfo {year} {2010})}\BibitemShut
  {NoStop}%
\bibitem [{\citenamefont {Santamore}\ \emph {et~al.}(2013)\citenamefont
  {Santamore}, \citenamefont {Lambert},\ and\ \citenamefont
  {Nori}}]{Santamore2013}%
  \BibitemOpen
  \bibfield  {author} {\bibinfo {author} {\bibfnamefont {D.}~\bibnamefont
  {Santamore}}, \bibinfo {author} {\bibfnamefont {N.}~\bibnamefont {Lambert}},
  \ and\ \bibinfo {author} {\bibfnamefont {F.}~\bibnamefont {Nori}},\
  }\href@noop {} {\bibfield  {journal} {\bibinfo  {journal} {PRB}\ }\textbf
  {\bibinfo {volume} {87}},\ \bibinfo {pages} {075422} (\bibinfo {year}
  {2013})}\BibitemShut {NoStop}%
\bibitem [{\citenamefont {Nishiguchi}\ and\ \citenamefont
  {Fujiwara}(2009)}]{Nishiguchi2009}%
  \BibitemOpen
  \bibfield  {author} {\bibinfo {author} {\bibfnamefont {K.}~\bibnamefont
  {Nishiguchi}}\ and\ \bibinfo {author} {\bibfnamefont {A.}~\bibnamefont
  {Fujiwara}},\ }\href@noop {} {\bibfield  {journal} {\bibinfo  {journal}
  {Nanotechnology}\ }\textbf {\bibinfo {volume} {20}} (\bibinfo {year}
  {2009})}\BibitemShut {NoStop}%
\end{thebibliography}

%

\end{document}